\documentclass[letterpaper]{article}

\usepackage{amsmath,amsfonts,amsthm}
\usepackage[bb=boondox]{mathalfa}
\usepackage[colorlinks=true,linkcolor=blue,citecolor=blue,urlcolor=blue]{hyperref}
\usepackage{authblk}

\numberwithin{equation}{section}

\newtheorem*{thm}{Theorem}

\theoremstyle{remark}
\newtheorem*{rmk}{Remark}

\let\Gamma\varGamma
\let\Delta\varDelta
\let\Theta\varTheta
\let\Lambda\varLambda
\let\Xi\varXi
\let\Pi\varPi
\let\Sigma\varSigma
\let\Upsilon\varUpsilon
\let\Phi\varPhi
\let\Psi\varPsi
\let\Omega\varOmega

\let\eps\varepsilon
\let\rho\varrho

\let\originalleft\left
\let\originalright\right
\renewcommand{\left}{\mathopen{}\mathclose\bgroup\originalleft}
\renewcommand{\right}{\aftergroup\egroup\originalright}

\newcommand{\dd}{\mathrm{d}}
\newcommand{\e}{\mathrm{e}}
\renewcommand{\i}{\mathrm{i}}
\newcommand{\R}{\mathbb{R}}
\newcommand{\Z}{\mathbb{Z}}
\newcommand{\N}{\mathbb{N}}

\newcommand{\C}{\Lambda}
\newcommand{\vol}{v}
\newcommand{\norm}[1]{\lvert #1 \rvert}
\newcommand{\Norm}[1]{\lVert #1 \rVert}
\newcommand{\inner}[2]{\langle #1, #2 \rangle}

\renewcommand{\P}{\Phi}
\newcommand{\Q}{\Phi^*}
\newcommand{\T}{\Theta}
\newcommand{\Pm}{\P/\T}
\newcommand{\Qm}{(\P/\T)^*}

\newcommand{\gm}{\gamma}
\newcommand{\M}{\eta}

\newcommand{\ii}{\alpha}

\newcommand{\wdot}{\kern1pt{\cdot}\kern1pt}

\begin{document}

\title{A Lower Bound on the Partition Function for a Strictly Neutral Charge-Symmetric System}
\author{Jeffrey P. Thompson\thanks{\href{mailto:jeff.thompson@utexas.edu}{jeff.thompson@utexas.edu}} }

\author{Isaac C. Sanchez\thanks{\href{mailto:sanchezn@che.utexas.edu}{sanchez@che.utexas.edu}}}

\affil{McKetta Department of Chemical Engineering, \\ 
The University of Texas at Austin, Austin, TX 78712, USA}

\date{\today}

\maketitle

\begin{abstract}
A lower bound on the grand partition function of a classical charge-symmetric system is adapted to the neutral grand canonical ensemble, in which the system is constrained to have zero total charge. This constraint permits us to consider two-body potentials that are only conditionally positive definite.
\end{abstract}

\section{Introduction}

Kennedy~\cite{Kennedy1982} obtained a lower bound on the grand canonical partition function of a  charge-symmetric system of classical particles interacting via a positive definite two-body potential. In the sine-Gordon representation, this lower bound is the Gaussian approximation obtained by retaining only the quadratic part of the cosine interaction. In this paper we establish an analogous lower bound on the \emph{neutral} grand canonical partition function, that is, the grand canonical partition function restricted to configurations with zero total charge.

The neutrality constraint allows us to consider interaction potentials that are only conditionally positive definite.\footnote{Cf. a remark in \cite[Example II.6.a]{Frohlich1978}.} Let $X$ be a nonempty set, and recall that a symmetric function $u\colon X \times X\to\R$ is called a \emph{conditionally positive definite kernel} if for any $n \ge 2$, $x_1,\dots,x_n\in X$, and $c_1,\dots,c_n\in\R$,
\begin{equation}
    \sum_{j,k=1}^n c_j c_k u(x_j,x_k) \ge 0 \quad \text{when } \sum_{j=1}^n c_j = 0.  \label{eq:cpd}
\end{equation}
In our context, $X$ is the configuration space of a single particle, and $u(x,y)$ is a two-body potential. Suppose $r$ is a metric on $X.$ Then $(x,y)\mapsto-r(x,y)$ and $(x,y)\mapsto -\log(1+r(x,y))$ are conditionally positive definite kernels. If $X$ is the torus $\R^d/a\Z^d$, then another example is
\begin{equation}
(x,y) \mapsto \frac{1}{a^d}\sum_{p \in X^*\setminus\{0\}}\frac{\e^{-t\norm{p}^2}}{\norm{p}^2} \, \e^{\i p \cdot (x-y)} \label{eq:pcp}
\end{equation}
with $X^*=2\pi a^{-1}\Z^d$ and $t>0$. From a physical point of view, these examples are Coulomb(-like) potentials with infrared singularities. 

An interesting consequence of strict neutrality is that the mean-field correlation length $\xi_0$ appearing in our lower bound depends on the volume.
(In the case of Coulomb interactions, $\xi_0$ is called the Debye length.) Suppose that $X$ is a Riemannian manifold and that the particles are confined to a bounded region $\C\subseteq X$ of Riemannian volume $\norm{\C}$. For a system of two species of particles with equal activities~$z$ and opposite charges~$\pm e$, one finds\footnote{A more general formula for $\xi_0$ is given by \eqref{eq:xi}.}
\begin{equation}
\xi_0 = \left[ 2 z \frac{I_1(2z\norm{\C})}{I_0(2z\norm{\C})} \beta e^2 \right]^{-1/2}. \label{eq:ld}
\end{equation}
Here $I_n$ is the $n$\textsuperscript{th} modified Bessel function, and $\beta$ is the inverse temperature.  The correlation length \eqref{eq:ld} should be compared with $(2z \beta e^2)^{-1/2},$ its counterpart in the ordinary grand canonical ensemble. 
The difference reflects the fact that the limit as $\beta \to 0$ of the neutral grand canonical particle density is 
\begin{equation}
2z\frac{I_1(2z\norm{\C})}{I_0(2z\norm{\C})} \label{eq:npd}
\end{equation}
instead of $2z$. Note that $(0,\infty) \to \R : x \mapsto I_1(x)/I_0(x)$ is a strictly increasing function with range $(0,1)$. Thus for given $z$ and $\norm{\C}$, \eqref{eq:npd} is smaller than $2z$, the difference vanishing for large $z\norm{\C}$. This effect of exact charge conservation on the density of an ideal gas of charged particles is termed \emph{canonical suppression} in high energy nuclear physics.\footnote{See e.g.~\cite{Fochler2006}. Note that in relativistic statistical thermodynamics, the term ``canonical'' is used to refer to the conservation of quantum numbers as opposed to particle numbers.}
In a mean-field treatment of interactions, this density suppression leads to enhancement of charge--charge correlations---that is,  suppression of screening.

\section{Definitions}
\label{sec:defs}

\subsection{The Model}

Our system consists of $s$ species of charged point particles moving in a space $X.$ We take $X$ to be either a complete connected Riemannian manifold $(X,g)$ or a lattice $l\Z^d \subset \R^d$. In the former case, $\vol$ denotes the volume measure induced by the metric $g$; in the latter, $\vol$ is the counting measure on $\Z^d$ multiplied by $l^d$, the volume of a lattice site. We sometimes use the notation $\norm{\,\cdot\,}=\vol(\wdot)$.

Let $z_\ii$ and $e_\ii$ denote the activity and charge, respectively, of the $\ii$\textsuperscript{th} species ($\ii=1,\dots,s$). The activity has dimensions of inverse volume. The charges are integer multiples of an elementary charge~$e$. 
We require charge symmetry. For the ordinary grand canonical ensemble, this condition is expressed by
\begin{equation*}
\sum_{\ii=1}^s e_\ii^{2n+1} z_\ii = 0  \quad\text{for all $n\in\N$}. 
\end{equation*}
For the neutral grand canonical ensemble, the corresponding condition is
\begin{equation}
\sum_{\ii=1}^s e_\ii^{2n+1}\e^{e_\ii c}z_\ii  = 0 \quad \text{for all $n\in \N$ for some $c\in\R$}.  \label{eq:cs}
\end{equation}
The difference stems from the fact that the neutral ensemble is invariant under the transformation $z_\ii \mapsto \e^{e_\ii c}z_\ii $. (See for instance \cite{Lieb1972}.) 

We define
$$
A_n = \big\{ (\ii_1,\dots,\ii_n) \in \{1,\dots,s\}^n : \textstyle{\sum_{j=1}^n e_{\ii_j} = 0 } \big\}.
$$
A \emph{neutral} configuration of $n$ particles is specified by a point
$$
((\ii_1,\dots,\ii_n),(x_1,\dots,x_n)) \in A_n \times X^n,
$$
$\ii_j$ and $x_j$ being the species and position, respectively, of the $j$\textsuperscript{th} particle.
The potential energy of a neutral configuration is given by
\begin{equation}
U_{\ii_1,\dots,\ii_n}(x_1,\dots,x_n) = \tfrac{1}{2}\sum_{j,k=1}^n e_{\ii_j}e_{\ii_k} u(x_j,x_k), \label{eq:u}
\end{equation}
where $u \colon X\times X\to \R$ is a conditionally positive definite kernel---see \eqref{eq:cpd}. Note that the diagonal terms are included in the sum. Since we assume only conditional positive definiteness, it is not correct in general to interpret $\tfrac{1}{2} e^2 u(x,x)$ as the self-energy of a charge $e$ at the point $x$. Moreover, the neutrality condition $(\ii_1,\dots,\ii_n)\in A_n$ permits  us to rewrite \eqref{eq:u} as
$$
 \sum_{1 \le j < k \le n} e_{\ii_j}e_{\ii_k} \left[u(x_j,x_k)-\tfrac{1}{2}u(x_j,x_j) -\tfrac{1}{2}u(x_k,x_k)\right], 
$$
where by construction the diagonal terms vanish. 

It is often convenient to subtract from \eqref{eq:u} a term
\begin{equation}
\tfrac{1}{2} u_0 \sum_{j=1}^n e_{\ii_j}^2, \label{eq:uref}
\end{equation}
where $u_0$ is a constant. For example, if $X$ is a periodic box $\R^d/a\Z^d$ of dimension $d > 2$ and $(x,y)\mapsto u(x,y)$ is given by \eqref{eq:pcp}, one might define
\begin{equation}
u_0 = \lim_{a \to \infty} \, \frac{1}{a^d} \sum_{\substack{p \in 2\pi a^{-1}\Z^d \\ p \ne 0}} \frac{\e^{-t\norm{p}^2}}{\norm{p}^2} =  \frac{2}{(d-2)(4\pi)^{d/2} \,t^{(d-2)/2}}. \label{eq:uref2}
\end{equation}
In this example, $\tfrac{1}{2}e^2 u_0$ is the self-energy of an isolated charge $e$ in the infinite-volume limit $a\to\infty$.
We absorb the term \eqref{eq:uref} into the activities, defining for each species $\ii$ the (dimensionless) measure
\begin{equation*}
\tilde{z}_\ii = \e^{(1/2)\beta e_\ii^2 u_0} z_\ii \vol. \label{eq:zt}
\end{equation*}
Note that charge symmetry is preserved---if the bare activities $z_\ii$ satisfy \eqref{eq:cs}, then the renormalized activities $\tilde{z}_\ii$ satisfy \eqref{eq:cs} with $0=(\text{trivial measure})$.

The neutral grand canonical partition function for a system of particles in a region $\C\subseteq X$ is given by
\begin{equation}
\Xi = \sum_{n=0}^\infty \frac{1}{n!} \sum_{(\ii_1,\dots,\ii_n) \in A_n} \int_{\C^n} \exp[-\beta U_{\ii_1,\dots,\ii_n}(x_1,\dots,x_n)] \prod_{j=1}^n \tilde{z}_{\ii_j}(\dd x_j). \label{eq:ngpf}
\end{equation}
The $n=0$ term is defined to be equal to $1$. We define the ideal-gas partition function $\Xi_0$ as the limit of $\Xi$ when $\beta\to 0$
 with the bare activities fixed:
\begin{equation}
\Xi_0 = \sum_{n=0}^\infty \frac{1}{n!} \sum_{(\ii_1,\dots,\ii_n) \in A_n} \prod_{j=1}^n (z_{\ii_j}\vol)(\C). \label{eq:igpf}
\end{equation}

\subsection{Charge Representation}

We proceed by recasting \eqref{eq:ngpf} into a more convenient form. This form is obtained by a change of variables $(\ii,x) \mapsto e_\ii \delta_x$, where  $e_\ii \delta_x$ is the \emph{charge distribution} of a particle of species $\ii$ located at $x$. (See for instance  \cite[Section~2]{Frohlich1981}.)

Let $\R^X$ be the space of all functions $\phi\colon X \to \R$ with the product topology. The dual space $(\R^X)^*$ consists of linear combinations of the projection maps ($\delta$ functions) $\delta_x \colon\phi \mapsto \phi(x)$, $x \in X.$  We introduce the notation
$$
\P = \R^X, \quad \Q = (\R^X)^*.
$$
In the present context, $\P$ is a space of scalar fields, and $\Q$ is a space of charge densities.
Let $\T\subset \P$ be the subspace of constant functions (i.e., constant fields).
Its annihilator
\begin{equation}
\T^\perp = \{ \rho \in \Q : \inner{\theta}{\rho} = 0 \text{ for all }\theta\in\T \} \label{eq:ann}
\end{equation}
is the subspace of \emph{neutral} charge densities.\footnote{In \eqref{eq:ann} and below, $\inner{\phi}{\rho}=\rho(\phi)$.} The space $\T^\perp$ may be identified with $\Qm$, the dual of the quotient space $\Pm$. To make this explicit, let $\pi$ be the quotient map that maps a field $\phi\in \P$ to its equivalence class
$$
\pi(\phi) = \{ \psi \in \P : \phi-\psi \in \T\} \in \Pm.
$$
The adjoint $\pi^* \colon \Qm\to\Q$ is an isomorphism onto its image $\T^\perp$.

Let $C^{\sim}$ be the symmetric bilinear form on $\Q$ defined by
$$
C^{\sim}(\delta_x,\delta_y) = u(x,y)
$$
and extension by linearity. Since $u$ is conditionally positive definite, we have
$C^{\sim}(\rho,\rho) \ge 0$ for all $\rho\in\T^\perp$. 
Hence the form $C$ on $\Qm$ defined by
\begin{equation}
C(\rho,\sigma) = C^{\sim}(\pi^*(\rho),\pi^*(\sigma)) \label{eq:c}
\end{equation}
is positive definite.
We interpret $C(\rho,\sigma)$ as the interaction potential between two neutral charge densities $\rho$ and $\sigma$. The potential energy of $n$ neutral charge densities $\rho_1,\dots,\rho_n$ is given by
\begin{equation}
U(\rho_1,\dots,\rho_n) = \tfrac{1}{2}\sum_{j,k=1}^n C(\rho_j,\rho_k). \label{eq:uc}
\end{equation}

We fix a basepoint $x_0$ in $X$ and define  $\delta'_x$, $x\in X$, by
$$
\pi^*(\delta'_x) = \delta_x - \delta_{x_0}.
$$
We shall think of a particle of species $\ii$ located at $x$ as having a charge distribution $e_\ii \delta'_x$. That is, each  charge $e_\ii$ is paired with a compensating charge $-e_\ii$ fixed at $x_0$.  (To generalize the term of \cite{Caillol1991}, we might call $\delta'_x$ a ``pseudocharge.'') The compensating charges sum to zero for neutral configurations:
$$
\sum_{j=1}^n e_{\ii_j} \pi^*(\delta'_{x_j}) = \sum_{j=1}^n e_{\ii_j} \delta_{x_j}
$$
when $(\ii_1,\dots,\ii_n)\in A_n$.

Let $f_1,\dots,f_s\colon X \to (\P/\T)^*$ be the maps defined by
$$
f_\ii(x) = e^{-1} e_\ii \delta'_x,
$$
where $e$ is the elementary charge. The $n$-particle potential \eqref{eq:u} is the pullback of the $n$-charge potential \eqref{eq:uc} along $ef_{\ii_1},\dots,ef_{\ii_n}$:
$$
U_{\ii_1,\dots,\ii_n}(x_1,\dots,x_n) = e^2 U(f_{\ii_1}(x_1),\dots,f_{\ii_n}(x_n)).
$$
We define a family of pushforward measures, $(\tilde{\lambda}_q)_{q\in\Z}$, by 
$$
\tilde{\lambda}_q(\wdot) = \sum_{\ii=1}^s \delta_{q,e^{-1}e_\ii} \tilde{z}_\ii(\C \cap f^{-1}_\ii(\wdot)). \label{eq:pf2}
$$
Here $\delta_{q,q'}$ is the Kronecker delta, and $f_\ii^{-1}$ denotes the preimage under $f_\ii$. Note that $\tilde{\lambda}_q \ne 0$ for only finitely many $q$. The bare counterpart of $\tilde{\lambda}_q$ is
$$
\lambda_q = \e^{-(1/2)\eps^2 q^2 u_0}  \tilde{\lambda}_q,
$$
where $\eps^2 = \beta e^2$.

We rewrite the partition function \eqref{eq:ngpf} as
\begin{equation}
\Xi = \sum_{n=0}^\infty \frac{1}{n!} \sum_{q_1,\dots,q_n \in \Z} \delta_{q_1+\dots+q_n,0} \\ \int_{(\Qm)^n} \e^{-\eps^2 U(\rho_1,\dots,\rho_n)} \prod_{j=1}^n \tilde{\lambda}_{q_j}(\dd \rho_j). \label{eq:ngpf2}
\end{equation}
The ideal-gas partition function  \eqref{eq:igpf} becomes
\begin{equation}
\Xi_0 = \sum_{n=0}^\infty \frac{1}{n!} \sum_{q_1,\dots,q_n \in \Z} \delta_{q_1+\dots+q_n,0} \prod_{j=1}^n \Norm{\lambda_{q_j}}, \label{eq:igpf2}
\end{equation}
where $\Norm{\lambda_q} = \lambda_q(\Qm)$.

\subsection{Sine-Gordon Representation}

We now carry out the ``Fourier transformation in the charge variables'' \cite{Frohlich1981}.

Let $\gm$ be the Gaussian measure on $\Pm$ with mean zero and covariance $C$.
We write the Boltzmann weight in \eqref{eq:ngpf2} in the form
$$
\delta_{q_1+\dots+q_n,0} \cdot \e^{-\eps^2 U(\rho_1,\dots,\rho_n)} = \int_0^{2\pi} \prod_{j=1}^n \e^{\i q_j \theta} \,\frac{\dd\theta}{2\pi} \cdot \int \prod_{j=1}^n \e^{\i \eps \inner{\phi}{\rho_j}} \gm(\dd\phi).
$$
Taking the sum over $n$ inside the $\theta$ and $\phi$ integrals gives
\begin{equation}
\Xi = \iint_0^{2\pi} \exp\!\Bigg[  \sum_{q\in\Z} \e^{\i q\theta} \int \e^{\i\eps\inner{\phi}{\rho}} \, \tilde{\lambda}_q(\dd\rho)  \Bigg] \frac{\dd\theta}{2\pi} \,\gm(\dd\phi). \label{eq:zsg} 
\end{equation}
The Fourier representation of \eqref{eq:igpf2} is
\begin{equation}
\Xi_0 = \int_0^{2\pi} \exp\!\Bigg[  \sum_{q\in\Z} \Norm{\lambda_q} \, \e^{\i q\theta} \Bigg] \frac{\dd\theta}{2\pi}. \label{eq:zsgi}
\end{equation}

So far we have not made use of the charge-symmetry condition \eqref{eq:cs}. 
This condition implies the relation
$$
\e^{- qc_0} \lambda_{ q}( Q) = \e^{qc_0} \lambda_{-q}(-Q),
$$
where $c_0$ is the real number satisfying
\begin{equation}
\sum_{q\in\Z} \e^{-qc_0} q \lambda_q  = 0, \label{eq:cc}
\end{equation}
and $-Q = \{\rho\in\Qm: -\rho\in Q\}$. We define 
\begin{gather*}
\tilde{\lambda}'_q = \e^{-q c_0} \tilde{\lambda}_q, \\
\lambda'_q = \e^{-q c_0} \lambda_q.
\end{gather*}
A complex translation $\theta \mapsto \theta + \i c_0$ in \eqref{eq:zsg} and \eqref{eq:zsgi} yields
\begin{gather}
\Xi = \iint_0^{2\pi} \exp\!\Bigg[  \sum_{q\in\Z} \int \cos(\eps\inner{\phi}{\rho} + q\theta) \,  \tilde{\lambda}'_q(\dd\rho)  \Bigg] \frac{\dd\theta}{2\pi} \,\gm(\dd\phi), \label{eq:zsg2} \\
\Xi_0 = \int_0^{2\pi} \exp\!\Bigg[  \sum_{q\in\Z} \Norm{\lambda'_q} \cos( q\theta) \Bigg] \frac{\dd\theta}{2\pi}. \nonumber 
\end{gather}

Let $\M$ be the probability measure on $\R/2\pi\Z$ defined by
\begin{equation*}
\M(\dd\theta) = (2\pi\Xi_0)^{-1}\exp\!\Bigg[  \sum_{q\in\Z} \Norm{\lambda'_q} \cos( q\theta) \Bigg]\dd\theta. \label{eq:vm}
\end{equation*}
We set
$$
V(\phi;\theta) = \sum_{q\in\Z} \Norm{\lambda'_q} \cos(q\theta) -\sum_{q\in\Z} \int \cos(\eps\inner{\phi}{\rho} + q\theta) \,\tilde{\lambda}'_q(\dd\rho)
$$
and write \eqref{eq:zsg2} as
$$
\Xi = \Xi_0 \iint_0^{2\pi} \e^{-V(\phi;\,\theta)} \, \M(\dd\theta) \,\gm(\dd\phi).
$$
This representation will be useful in the following section.

\section{Results}
\label{sec:results}

In this section, we obtain a lower bound analogous to that of \cite{Kennedy1982} on the neutral grand canonical partition function $\Xi.$
Let us write
$$
\Xi = \Xi_0 \int \e^{-V(\phi)} \, \gm(\dd\phi),
$$
where
$$
\e^{-V(\phi)} = \int_0^{2\pi} \e^{-V(\phi;\,\theta)} \, \M(\dd\theta).
$$
Expanding $V(\phi)$ to second order in $\eps$,\footnote{Recall that $\tilde{\lambda}'_q = \exp(\tfrac{1}{2}\eps^2 q^2 u_0 ) \lambda'_q.$} we obtain
\begin{equation}
V_2(\phi) = \tfrac{1}{2} \eps^2 \sum_{q\in\Z} \hat{\M}_q  \int \inner{\phi}{\rho}^2 \, \lambda'_q(\dd\rho) - \tfrac{1}{2} \eps^2 u_0 \sum_{q\in\Z} \hat{\M}_q \Norm{\lambda'_q} q^2, \label{eq:v2}
\end{equation}
where $\hat{\M}_q = \int_0^{2\pi} \e^{\i q\theta} \, \M(\dd\theta)$ are the Fourier coefficients of $\M$. Note that
\begin{equation}
\hat{\M}_{\pm q} = \frac{1}{\Xi_0}\sum_{n=0}^\infty \frac{1}{n!} \sum_{q_1,\dots,q_n \in \Z} \delta_{q_1+\dots+q_n,\mp q} \prod_{j=1}^n \Norm{\lambda'_{q_j}} \ge 0. \label{eq:xx}
\end{equation}
Our main result is the following theorem.

\begin{thm}
\begin{equation}
\Xi \ge \Xi_0 \int \e^{-V_2(\phi)} \, \gm(\dd\phi) . \label{eq:lb}
\end{equation}
\end{thm}

\begin{rmk}
We  prove \eqref{eq:lb} by using Jensen's inequality. This method is suggested by the final remark of \cite{Kennedy1982}. 
\end{rmk}

\begin{proof}
Define the normalized Gaussian measure
$$
\gm_2(\dd\phi) = \Xi_2^{-1} \Xi_0 \,\e^{-V_2(\phi)}\, \gm(\dd\phi), 
$$
where $\Xi_2$ is the right-hand side of \eqref{eq:lb}, 
and write $\Xi$ as
$$
\Xi = \Xi_2 \iint_0^{2\pi} \e^{V_2(\phi)-V(\phi;\,\theta)} \,\M(\dd\theta) \, \gm_2(\dd\phi).
$$
By Jensen's inequality,
\begin{equation}
\Xi \ge \Xi_2 \, \e^{E} \label{eq:jensen}
\end{equation}
with
$$
E = \int V_2(\phi) \, \gm_2(\dd\phi) - \iint_{0}^{2\pi} V(\phi;\theta) \, \M(\dd\theta) \, \gm_2(\dd\phi).
$$
Carrying out the $\phi$ and $\theta$ integrations gives
$$
E = \sum_{q\in\Z} \hat{\M}_q \int \big[ \e^{c_q(\rho)} - 1 - c_q(\rho) \big]  \lambda'_q(\dd\rho),
$$
where
$$
c_q(\rho) = \tfrac{1}{2}\eps^2  q^2 u_0 - \tfrac{1}{2}\eps^2 \int \inner{\phi}{\rho}^2 \, \gm_2(\dd\phi).
$$
Recalling \eqref{eq:xx} and noting that $e^{c}-1-c \ge 0$ for all $c\in\R$, we see that $E \ge 0$. Thus the inequality $\Xi\ge \Xi_2$ follows from \eqref{eq:jensen}.
\end{proof}

As mentioned in the Introduction, the lower bound $\Xi_2$ involves a volume-dependent correlation length $\xi_0$. To see this, let us write \eqref{eq:v2} as
$$
V_2(\phi) = \tfrac{1}{2} \xi_0^{-2} \int_\C \inner{\phi}{\delta'_x}^2 \, \vol(\dd x) - \tfrac{1}{2} \xi_0^{-2} u_0 \norm{\C},
$$
where 
\begin{equation}
\xi_0 = \left( \beta \sum_{\ii=1}^s e^2_\ii z'_\ii \hat{\M}_{e^{-1}e_\ii} \right)^{-1/2} \label{eq:xi}
\end{equation}
with $$
z'_\ii = \e^{-e^{-1}e_\ii c_0} z_\ii.
$$
(The constant $c_0$ satisfies \eqref{eq:cc}.) The  correlation length $\xi_0$ 
 depends on the volume through the factors
$$
\hat{\M}_{e^{-1}e_\ii} = \frac{e}{2\pi\Xi_0}\int_0^{2\pi/e} \exp\!\Bigg[ \norm{\C} \sum_{\ii=1}^s z'_\ii \cos(e_\ii \theta) \Bigg] \e^{\i e_\ii \theta} \, \dd\theta.
$$
Note that $\lim_{\norm{\C}\to\infty}\hat{\M}_{e^{-1}e_\ii}=1$, so 
$$
\xi_0 \to \left( \beta \sum_{\ii=1}^s e^2_\ii z'_\ii  \right)^{-1/2}
$$
in the infinite-volume limit.

\appendix

\section*{Appendix}

The Gaussian approximation $\Xi_2$ can be computed explicitly. As an illustration, we give in this Appendix an explicit formula for the case of a Coulomb system in a torus
$$
\C=X= \R^d/a\Z^d.
$$
The interaction potential is assumed to be given by \eqref{eq:pcp}; that is,
$$
u(x,y) = \frac{1}{\norm{\C}}\sum_{p \in \C^* \setminus \{0\}} \hat{u}_p \, \e^{\i p \cdot(x-y)},
$$
where $\norm{\C} = a^d$, $\C^*=2 \pi a^{-1} \Z^d$, and 
$$
\hat{u}_p = \frac{\e^{-t \norm{p}^2}}{\norm{p}^2}.
$$
The parameter $t > 0$ is an ultraviolet cutoff. (For $d=1$ we may let $t = 0$.) We assume the term $u_0=u_0(t)$ introduced in \eqref{eq:uref} is  chosen so that the energy
$$
E_0 = \tfrac{1}{2}\lim_{t\to 0} \Bigg[{ u_0(t) - \frac{1}{\norm{\C}}\sum_{p\in\C^*\setminus\{0\}} \frac{\e^{-t\norm{p}^2}}{\norm{p}^2} } \Bigg]
$$
is finite.

For $\phi \in \Pm$ and $p\in\C^*\setminus\{0\}$, let
$$
\hat{\phi}_p = \int_\C \inner{\phi}{\delta'_x} \, \e^{-\i p \cdot x} \, \vol(\dd x).
$$
(Recall that $\pi^*(\delta'_x) = \delta_x - \delta_{x_0}$,  and note that $\hat{\phi}_p$ is independent of the choice of basepoint $x_0$.) The Fourier components $\hat{\rho}_p$ of a charge distribution $\rho\in\Qm$ satisfy
$$
\frac{1}{\norm{\C}} \sum_{p \in \C^*\setminus\{0\}} \hat{\phi}_p \,\overline{\hat{\rho}_p}  = \inner{\phi}{\rho}
$$
for any $\phi\in\Pm$. Explicitly, if $\rho=\sum_{j=1}^n c_j \delta'_{x_j}$ with $\sum_{j=1}^n c_j = 0$, then
$$
\hat{\rho}_p = \sum_{j=1}^n c_j \e^{-\i p \cdot x_j}.
$$
We have
$$
V_2(\phi) = \tfrac{1}{2} \xi_0^{-2} \norm{\C}^{-1} \sum_{p\in\C^*\setminus\{0\}} \norm{\hat{\phi}_p}^2 - \tfrac{1}{2} \xi_0^{-2} u_0 \norm{\C}
$$
and 
$$
\int \e^{\i\inner{\phi}{\rho}} \, \gm(\dd\phi) = \exp\!\Bigg[{- \tfrac{1}{2}\norm{\C}^{-1} \sum_{p \in \C^*\setminus\{0\}} \hat{u}_p \norm{\hat{\rho}_p}^2 }\Bigg].
$$
A standard calculation gives
\begin{align*}
\Xi_2 &= \Xi_0 \int \e^{-V_2(\phi)} \, \gm(\dd\phi) \\
&= \Xi_0 \exp\!\Bigg[ \tfrac{1}{2} \xi_0^{-2} u_0 \norm{\C} - \tfrac{1}{2} \sum_{p \in \C^*\setminus\{0\}} \log \left( 1 + \xi_0^{-2}\hat{u}_p \right)\Bigg].
\end{align*}

In dimensions $d < 4$, we can remove the ultraviolet cutoff in $\Xi_2$ by letting $t$ go to zero. The result is the  Debye--H\"{u}ckel approximation
$$
\lim_{t \to 0} \Xi_2 = \Xi_0  \exp\!\Bigg\{{ \xi_0^{-2} E_0 \norm{\C}- \tfrac{1}{2} \sum_{p \in \C^*\setminus\{0\}} \left[\log \left( 1 + \xi_0^{-2}\norm{p}^{-2} \right)-\xi_0^{-2}\norm{p}^{-2}\right]}\Bigg\}.
$$

\bibliographystyle{abbrv}
\bibliography{library}

\end{document}